\documentclass[12pt]{article}

\usepackage[numbers,sort&compress]{natbib}
\usepackage{graphicx}
\usepackage{amsmath}
\usepackage{amssymb}
\usepackage{hyperref}

\def\be{\begin{equation}}
\def\ee{\end{equation}}
\def\bsp{\be\begin{split}}

\def\wt{\widetilde}

\def\G{\Gamma}
\def\D{\Delta}

\def\a{\alpha}
\def\b{\beta}

\def\d{\delta}
\def\e{\epsilon}
\def\m{\mu}

\def\n{\nu}
\def\s{\sigma}
\def\r{\rho}
\def\l{\lambda}

\def\O{\Omega}

\def\vt{\vartheta}

\def\p{\partial}

\def\bR {\mathbb{R}}

\makeatletter

\newcommand{\Rmnum}[1]{\expandafter\@slowromancap\romannumeral #1@}
\makeatother

\newcommand{\beq}{\begin{equation}}
\newcommand{\eeq}{\end{equation}}
\newcommand{\bea}{\begin{eqnarray}}
\newcommand{\eea}{\end{eqnarray}}
\newcommand{\rf}[1]{(\ref{#1})}
\newcommand{\Urm}{\mathrm{U}}
\renewcommand{\title}[1]{\vbox{\center\LARGE{#1}}\vspace{5mm}}
\renewcommand{\author}[1]{\vbox{\center\large{#1}}\vspace{5mm}}
\newcommand{\address}[1]{\vbox{\center\em#1}}
\newcommand{\email}[1]{\vbox{\center\tt#1}\vspace{5mm}}

\newcommand{\Tr}{\mathrm{Tr}}

\newcommand{\Ncal}{\mathcal{N}}

\newcommand{\Cset}{{\,\,{{{^{_{\pmb{\mid}}}}\kern-.47em{\mathrm C}}}}}

\newcommand{\diff}{\mathrm{d}}
\newcommand{\half}{\frac12}

\newcommand{\comment}[1]{}
\begin{document}
\bibliographystyle{utphys}
\newpage
\setcounter{page}{1}
\pagenumbering{arabic}
\renewcommand{\thefootnote}{\arabic{footnote}}
\setcounter{footnote}{0}

\begin{titlepage}
\title{\vspace{1.0in} {\bf Holographic three-point functions of giant gravitons}}
 
\author{A.\ Bissi$^{a,b}$, C.\ Kristjansen$^a$, D.\ Young$^a$ and K.\ Zoubos$^a$}

\address{$^a$Niels Bohr Institute, Blegdamsvej 17, DK-2100 Copenhagen, Denmark \\
$^b$Niels Bohr International Academy, Niels Bohr Institute, Blegdamsvej 17, DK-2100
Copenhagen, Denmark}

\email{bissi, kristjan, dyoung, kzoubos@nbi.dk}

\abstract{Working within the AdS/CFT correspondence we calculate the
three-point function of two giant gravitons and one pointlike graviton
using methods of semiclassical string theory and considering both the
case where the giant gravitons wrap an $S^3\subset S^5$ and the case
where the giant gravitons wrap an $S^3\subset AdS_5$. We likewise
calculate the correlation function in ${\cal N}=4$ SYM using
two Schur polynomials and a single trace chiral primary. We find that
the gauge and string theory results have structural similarities but
do not match perfectly, and interpret this
in terms of the Schur polynomials' inability to interpolate between
dual giant and pointlike gravitons. }
\end{titlepage}

\section{Introduction}

\indent The integrable structures underlying the $\Ncal=4$
Super-Yang-Mills (SYM) theory are a continuing source of fascination,
in large part due to the promise they hold of leading to the complete
solution of this non-trivial four-dimensional gauge theory. Since
their appearance in \cite{Minahan:2002ve,Beisert:2003tq}, the main focus of
investigation has been the \emph{spectral problem}, i.e.  the question
of obtaining the exact spectrum of anomalous dimensions of
gauge-invariant operators of the theory. For a review of the state of
the art and current challenges in this field, we refer to the
collection of articles in \cite{Beisert:2010jr}. The spectrum,
however, is only part of the information required to (at least in
principle) solve the theory, the other essential ingredient being the
set of all three-point functions between the fundamental operators. It
 thus natural to look beyond the spectrum and ask whether
integrability plays any role in specifying the correlation functions
of the theory. Answering this question is also likely to be crucial in
attempts to probe integrability beyond the planar limit, which
involves considering interacting strings (and three-point vertices
thereof) in the dual description.\footnote{See the review
\cite{Kristjansen:2010kg} for a recent discussion of these issues.}
 
 Understanding the role of integrability in the calculation of
$\Ncal=4$ SYM correlation functions has recently been receiving a
growing amount of attention. On the gauge side of the AdS/CFT
correspondence, and following earlier work in \cite{Okuyama:2004bd,Roiban:2004va,Alday:2005nd},
it was shown in \cite{Escobedo:2010xs} that the
one-loop Bethe equations can be used to simplify the calculation of
tree-level three-point functions of certain non-protected operators.

On the gravity side, the computation of two-point functions of
spinning strings was recently discussed in \cite{Janik:2010gc}, as
well as in \cite{Buchbinder:2010vw} using the formalism of
semiclassical vertex operators \cite{Tseytlin:2003ac,Buchbinder:2010gg}. However, the
extension to general semiclassical three-point functions appears to
rely on finding the precise geometric solution interpolating between
the insertion points of the corresponding operators on the boundary of
AdS, which seems to lie beyond current capabilities. As argued in
\cite{Zarembo:2010rr,Costa:2010rz}, this problem can be avoided
provided one restricts to correlation functions involving only two
``heavy'' operators (for which the classical string trajectory is
known), while treating the remaining ``light''operators as a
perturbation.  With this simplifying assumption, several cases of
three- and higher-order correlation functions, involving different
types of semiclassical strings and various choices for the light
operators, have been considered in the literature
\cite{Roiban:2010fe,Hernandez:2010tg,Ryang:2010bn,Arnaudov:2010kk,Georgiou:2010an,Russo:2010bt,Park:2010vs,Buchbinder:2010ek}.

 The semiclassical string solutions discussed above correspond to
single-trace gauge theory operators with large quantum numbers, which
can also be described as spin chains with certain amounts of
excitations. Although the dimension of these states is much larger than
1 in order to justify the semiclassical approximation (in particular,
the dimension is $\sim \sqrt{\lambda}$ with $\lambda$ the 't Hooft
coupling), it is also necessarily much smaller than the rank of the
gauge group $N$. However, the gauge theory also contains operators
whose dimension scales as $N$ in the large-$N$ limit. A class of such
operators was identified in \cite{Balasubramanian:2001nh} with the
\emph{giant gravitons} of \cite{McGreevy:2000cw}, which are D3-branes
wrapping an $S^3$ in the $S^5$ (their radius stabilised by the
presence of five-form flux in the geometry), spinning along a circle
in the $S^5$ and located at the centre of $AdS_5$. The mapping of
these giant gravitons, as well as the dual giant gravitons wrapping an
$S^3$ in $AdS_5$ constructed in \cite{Hashimoto:2000zp}, to the
gauge theory was put in a more general context in the work of
\cite{Corley:2001zk}: they correspond to \emph{Schur polynomial}
operators, which are specific combinations of traces of one of the
$\Ncal=4$ SYM scalars forming an orthogonal basis for any
$N$.\footnote{See section 2 for the precise definition of these
operators.}

 Each Schur polynomial is labeled by a Young tableau corresponding to
a specific representation of $\Urm(N)$.  As argued in
\cite{Corley:2001zk}, $S^5$ giant gravitons of dimension $k\leq N$
are mapped to the \emph{antisymmetric} representation with $k$ boxes,
while the dual $AdS_5$ ones map to the \emph{symmetric} representation
with $k$ boxes.  The fact that the dimension of the antisymmetric
representation is bounded from above by $N$ corresponds to the fact
that the angular momentum of the $S^5$ graviton has an upper bound:
its radius increases with angular momentum but cannot become larger
than that of the $S^5$. Giant gravitons which saturate this bound
are called \emph{maximal}.

 Given that giant gravitons are heavy, semiclassical objects, one can
ask whether the approach of \cite{Zarembo:2010rr} can be
applied to correlation functions of giant gravitons.  A calculation
involving giant gravitons was recently performed in \cite{Bak:2011yy},
which considered correlation functions of two $S^5$ giant gravitons
with open strings attached (as the heavy states), with the light state
being dual to a chiral primary operator.  This work computed two-point
functions for both maximal and non-maximal giant gravitons, but
three-point functions only in the maximal case, where the giant
gravitons were essentially inert (their role being that of providing
the open string endpoints).

 In this work we consider a different type of correlation function of
two giant graviton operators with a chiral primary.  On the string
side, we will take as our heavy states either $AdS_5$ or $S^5$
giant gravitons (not necessarily maximal). The light state will be an
AdS scalar field of dimension $\D=J\ll\sqrt{N}$ dual to a chiral
primary. On the gauge theory side, the operators corresponding to the
giant gravitons are Schur polynomials with dimension $k$ of order
$N$, in the large-$N$ limit. We compute the three-point function of
these operators with a chiral primary operator, using established
gauge theory techniques. This quantity is expected to be protected
from quantum corrections, owing to the shared 1/2 BPS supersymmetry of
the operators, however we do not find perfect agreement with the string
theoretic results.  In the concluding section we discuss in detail
this discrepancy which we believe hinges on the inability of the Schur
polynomials to interpolate between giant and pointlike gravitons.

The plan of the paper is as follows: In the next section we introduce
the operators that we will consider on the gauge theory side and
compute the three-point functions in question using Schur polynomial
techniques.  Then, in section \ref{stringside} we describe the string
theory computation of the same quantities using the approach of
\cite{Zarembo:2010rr}. In the concluding section we
compare the two results and discuss open problems and directions for
future work. We have also included an appendix containing a simple
computation of the holographic $S^5$ giant graviton two-point function.

\section{Three-point functions from gauge theory}

In this section we will compute three-point functions involving two types of half-BPS
operators: single trace chiral primaries (which in the following we will simply 
call ``chiral primaries'') and Schur polynomial operators. Three-point functions
of such operators (all built up of the same $\Ncal=4$ SYM chiral field $Z$) 
are expected to be protected to all orders \cite{D'Hoker:2002aw} and 
can thus be exactly calculated in the gauge theory.

\subsection{Single trace chiral primaries}

 Let us consider single-trace operators built from a single complex scalar field $Z$, i.e.
\begin{equation}
{\cal O}^J=\Tr Z^J, \label{singletrace}
\end{equation}
These operators are dual to point-like strings moving along an equator of
$S^5$ with angular momentum $J$.
Their two- and three point  functions are protected
and can be calculated exactly, see for instance \cite{Kristjansen:2002bb}, and read
\bea
\langle \Tr Z^J \, \Tr\bar{Z}^J \rangle &=&
 \frac{1}{J+1}
\left\{\frac{\Gamma(N+J+1)}{\Gamma(N)}-\frac{\Gamma(N+1)}{\Gamma(N-J)}
\right\} \label{Tracenorm}  \\
&=& \label{Trnorm}
J\, N^J\left\{ 1+
\left( \begin{array}{c} J+1 \\ 4\end{array} \right)
\frac{1}{N^2}
+\ldots \right\}.
\eea
\bea
\lefteqn{\langle \Tr Z^J \,\Tr Z^K \Tr\bar{Z}^{J+K} \rangle}\nonumber \\ &=&
 \frac{1}{J+K+1}
\left\{\frac{\Gamma(N+J+K+1)}{\Gamma(N)}-\frac{\Gamma(N+J+1)}{\Gamma(N-K)}
\right.\nonumber  \\
&& \left.
+\frac{\Gamma(N+1)}{\Gamma(N-J-K)}-\frac{\Gamma(N+K+1)}{\Gamma(N-J)}
\right\} \nonumber \\
&=& 
 N^{J+K-1} J\, K\, (J+K) \,\\
&&\times\left\{ 1+ \frac{1}{3!N^2}
\left( \begin{array}{c} K+J-1 \\ 2\end{array} \right)
\left[\left( \begin{array}{c} K \\ 2\end{array} \right)
+\left( \begin{array}{c} J \\ 2\end{array} \right)-1
\right]
+\ldots \right\}.\nonumber
\eea 
Here we have left out the trivial dependence on space-time coordinates
and the 't Hooft coupling constant. Hence we get for the CFT structure
constant
\bea
{\cal C}_{J,K,K+J}&\equiv& \frac{\langle{\cal O}^J {\cal O}^K 
\bar{{\cal O}}^{J+K}\rangle}
{\sqrt{\langle{\cal O}^J\bar{{\cal O}}^J\rangle
\langle{\cal O}^K\bar{{\cal O}}^K\rangle
\langle{\cal O}^{J+K}\bar{{\cal O}}^{J+K}\rangle
}} \\
&=&
\frac{1}{N} \sqrt{ J\,K\,(J+K)}
\left[1+{\cal O}\left(\frac{1}{N^2}\right)\right].
\label{3ptchiral}
\eea
This is the well known expression for the three-point function of three
chiral primaries of the type given in eqn.~\rf{singletrace}. 
This object can also
be viewed as a two point function of a single trace operator and a double
trace operator (since the contractions needed are the same in both cases)
and we notice the well-known fact
 that single and multi-trace operators are orthogonal to the
leading order in $\frac{1}{N}$ provided $J$ is not too big. For large 
values of $J$ single trace operators mix with multi trace operators and
a more convenient basis is the basis of Schur polynomials $\chi_R(Z)$
described below~\cite{Corley:2001zk}.

\subsection{Schur polynomials}
The Schur polynomial $\chi_R(Z)$ of a complex matrix $Z$ is defined as
\beq
\chi_{R_n}(Z)=\frac{1}{n!}
\sum_{\sigma\in S_n} \chi_{R_n}(\sigma) \,Z_{i_1}^{i_{\sigma(1)}}
\ldots Z_{i_n}^{i_{\sigma(n)}}.
\eeq
Here $R_n$ denotes an irreducible representation of $U(N)$ described
in terms of a Young tableau with $n$ boxes. The sum is over all elements of
the symmetric group $S_n$
 and $\chi_{R_n}(\sigma)$ is the character of the
element $\sigma$ in the representation $R_n$. Notice that there is no
limit in which the Schur polynomial reduces to a chiral primary 
operator.\footnote{The Schur polynomials
$\chi_{R_n}(Z)$ have the general structure
\beq
\chi_{R_n}(Z)=c_{0,n}\Tr Z^n+c_{1,n}\Tr Z \Tr Z^{n-1}+\ldots +
c_{n,n}\,(\Tr Z)^n,
\eeq
where the $c$'s are constants independent of $N$.}
Schur polynomials are again 1/2-BPS operators with protected two- and 
three point functions. These correlation functions have been calculated
exactly and read~\cite{Corley:2001zk}
\bea
\langle \chi_{R}(Z)\chi_{S}(\bar{Z})\rangle &=& \delta_{R,S} 
\prod_{i.j\in R}(N-i+j) \label{normgiant},\\
\langle \chi_{R}(Z)\chi_{S}(Z) \chi_{T}(\bar{Z})\rangle
&=&g(R,S;T)\, \prod_{i,j\in T}(N-i+j) \label{Littlewood},
\eea
where $g(R,S,T)$ is the Littlewood-Richardson coefficient which counts the
multiplicity with which the representation $T$ appears in the tensor product
of the representations $R$ and $S$. Furthermore, the product $\prod_{i,j\in R}$
goes over all
boxes of the Young tableau of the representation $R$
with $i$ denoting the row number and $j$ the
column number.
Hence the Schur polynomials provide an orthogonal basis of
operators. The string theory duals of Schur polynomials are collections of
giant gravitons,
i.e. D3-branes which wrap an $S^3$ of either $S^5$ or 
$AdS_5$~\cite{Balasubramanian:2001nh,Corley:2001zk}.
The cleanest examples are the Schur polynomials of the
symmetric and  the antisymmetric representations. When the number of 
boxes, $k$, in the Young tableau of the representation is large 
(i.e.\ $k\sim {\cal O}(N)$, with $N\rightarrow \infty$), 
the Schur polynomial of the symmetric representation is dual to a single 
giant graviton moving on $S^5$ with angular momentum
$k$ and
wrapping an $S^3\subset AdS_5$. For the antisymmetric
case the giant graviton instead wraps an $S^3\subset S^5$~\cite{Corley:2001zk}.

Let us denote the Schur polynomial for the symmetric representation with
$k$ boxes as $\chi_k^S(Z)$ and the Schur polynomial for the antisymmetric 
representation with $k$ boxes as $\chi_k^A(Z)$. Then
 we find for the corresponding
two and three-point functions
\bea
\langle \chi_k^S(\bar{Z}) \chi_{k}^S(Z) \rangle&=&
\prod_{j=1}^k (N-1+j), \label{schursym2} \\
 \langle \chi_k^A(\bar{Z}) \chi_{k}^A(Z) \rangle&=& 
\prod_{i=1}^k (N-i+1), \label{schurantisym2}\\
\langle \chi_k^S(\bar{Z}) \chi_{k-J}^S(Z) \chi_J^S (Z) \rangle
&=&\prod_{j=1}^k (N-1+j), \label{schursym3}\\
\langle \chi_k^A(\bar{Z}) \chi_{k-J}^A(Z) \chi_J^A(Z) \rangle
&=&\prod_{i=1}^k (N-i+1), \label{schurantisym3}
\eea 
since for these cases $g(R,S;T)=1$. Notice that for the antisymmetric case
we have that $k\leq N$ while in the symmetric case $k$ is
unbounded.

\subsection{Two Schur polynomials and one single trace operator}

Recently, it has been understood from the string theory side 
how to calculate by 
semiclassical methods three-point functions which involve two massive
string states and one light one dual to a chiral primary
operator of the type $\Tr Z^J$~\cite{Janik:2010gc,Zarembo:2010rr}. 
In the present 
paper we will study the case where the two heavy operators are giant
gravitons. In the field theory
language these three-point functions can be calculated exactly using 
the results of the sections above. The properly normalized three-point
functions are
\beq
C_{k,k-J,J}^S \equiv
\frac{\langle \chi_k^S(\bar{Z}) \chi_{k-J}^S(Z) \Tr Z^J \rangle}
{\sqrt{\langle \chi_k^S(\bar{Z}) \chi_{k}^S(Z)\rangle 
\langle \chi_{k-J}^S(\bar{Z}) \chi_{k-J}^S(Z)\rangle
\langle \Tr \bar{Z}^{J} \Tr Z^{J}\rangle
}},\label{CijkS}
\eeq
and similarly for $C_{k,k-J,J}^A$. We have already calculated the relevant
norms above, cf.\ eqns.\ \rf{Tracenorm} and~\rf{normgiant}. To calculate the expectation value
in the numerator 
we expand $\Tr \, Z^J$ in the
basis of Schur polynomials.
Noting that by definition 
\beq
\Tr Z^J = \Tr(\sigma_0 Z),
\eeq
where $\sigma_0$ is the cyclic permutation we have
\beq
\Tr\, Z^J =\sum_{R_J} \chi_{R_J}(\sigma_0) \chi_{R_J}(Z),
\eeq
where the sum goes over all possible irreducible representations $R_J$ 
corresponding
to Young tableaux with $J$ boxes, see e.g.\ \cite{Corley:2002mj}. 
Inserting the sum instead of $\Tr Z^J$
in the expectation values 
$\langle \chi_{k}^A(\bar{Z}) \chi_{k-J}^A(Z)\, \Tr \,Z^J\rangle$ and
$\langle \chi_{k}^S(\bar{Z}) \chi_{k-J}^S(Z)\, \Tr \,Z^J\rangle$ it is clear 
from~\rf{Littlewood} that only the completely antisymmetric representation
contributes
in the former case and only 
the completely symmetric representation in the latter. 
The character $\chi_{R_J}(\sigma_0)$ can be written down
in closed  form for  hook diagrams,
i.e.\ Young diagrams for which only the first row can have more than
one box. Denoting the number of boxes in the first row of the hook diagram
as $J-m$ it holds that $\chi_{R_J}^{hook}(\sigma_0)=(-1)^m$. Hence for the cases of
interest to us we have
\beq
\chi_J^S(\sigma_0)=1, \hspace{0.5cm} \chi_J^A(\sigma_0)= (-1)^{J-1}.
\eeq
 This implies\footnote{The $(-1)^J$ part of the prefactor in the antisymmetric case could be removed since
one can equally well define the gauge theory dual of the antisymmetric giant graviton with angular momentum $k$ to
be $(-1)^k\chi_k^A(Z)$. However, in the following we will follow the usual definition in the Schur operator literature and keep the alternating sign.}
\bea
\langle \chi_{k}^S(\bar{Z}) \chi_{k-J}^S(Z)\, \Tr \,Z^J\rangle &=& 
\prod_{j=1}^k(N-1+j),\\
\langle \chi_{k}^A(\bar{Z}) \chi_{k-J}^A(Z)\, \Tr \,Z^J\rangle &=& (-1)^{J-1}
\prod_{i=1}^k (N-i+1).
\eea
Dividing with the relevant norms we hence find the structure constants
\beq
C_{k,k-J,J}^S= 
\frac{\sqrt{\prod_{p=k-J+1}^k(N+p-1)}}
{\sqrt{JN^J (1+c(J)\,
\frac{1}{N^2}+\ldots)}},
\eeq
\beq
C_{k,k-J,J}^A= 
(-1)^{(J-1)}\frac{\sqrt{\prod_{p=k-J+1}^k(N-p+1)}}
{\sqrt{JN^J(1+c(J)\,\frac{1}{N^2}+\ldots)}},
\eeq
where the quantities in the denominators are nothing but 
 $\sqrt{\langle \Tr \bar{Z}^J \Tr Z^J\rangle}$ which is given exactly
in equation~\rf{Tracenorm}. In other words we have exact expressions
for $C_{k,k-J,J}^S$ and $C_{k,k-J,J}^A$. Now, we are interested in
the situation where the Schur polynomials correspond to large Young
tableaux and where the chiral primary is a small operator, i.e. the
limit
\beq
N\rightarrow \infty,\hspace{0.7cm} k\rightarrow \infty,
\hspace{0.7cm}\frac{k}{N}\,\,\,\,\mbox{finite}, \hspace{0.7cm}
J\ll k,
\eeq
and in particular $J\ll \sqrt{N}$. In this limit we find for the 
structure constants
\bea
C_{k,k-J,J}^S &= & \frac{1}{\sqrt{J}}\left(1+\frac{k}{N}\right)^{J/2}, 
\label{CS}\\
C_{k,k-J,J}^A & =&  (-1)^{(J-1)}\frac{1}{\sqrt{J}}\left(1-\frac{k}{N}\right)^{J/2}.
\label{CA}
\eea
Notice that this result does not reduce to the chiral primary result
in any limit (in accordance with the fact that a chiral primary operator
can not be obtained as a limit of a single Schur polynomial). Furthermore,
we note that for the antisymmetric representation we have the
constraint $k\leq N$ while for the symmetric case $k$ is unbounded.

\section{Three point function from string theory \label{stringside}}

In this section we will calculate the three-point function structure
constants considered in the previous sections using the AdS/CFT
dictionary put forth in \cite{Janik:2010gc,Zarembo:2010rr}. We work
under the assumption that the holographic two-point function of the
giant gravitons are given by the D-brane solutions
\cite{McGreevy:2000cw,Grisaru:2000zn} continued to the Euclidean
Poincar\'{e} patch, in the same way that semiclassical spinning
strings were argued to represent the two-point functions of the
associated operators in \cite{Janik:2010gc,Zarembo:2010rr}. The
calculation proceeds by varying the Euclidean D-brane actions in
accordance with the supergravity fluctuations corresponding to the
small operator in the desired three-point function, and then
evaluating those fluctuations on the Wick-rotated giant graviton
solutions, described in the Poincar\'{e} patch.\footnote{Such fluctuation
calculations for D-branes were first performed in \cite{Giombi:2006de} in 
the context of Wilson loops in higher representations.}

\subsection{Giant graviton on $S^5$}
\label{sec:s5}

We begin by reviewing the giant graviton \cite{McGreevy:2000cw} with
worldvolume $\bR (\subset AdS_5) \times S^3 (\subset S^5)$. We begin
in Lorentzian signature $(-,+,\ldots,+)$. The metric of $AdS_5\times
S^5$ can be taken as
\be\label{global}
ds^2  = -\cosh^2 \r\, dt^2 + d\r^2 + \sinh^2\r\, d\wt\O_3^2
+ d\theta^2 + \sin^2\theta\, d\phi^2 + \cos^2\theta \,d\O_3^2.
\ee
The action for the D3-brane is (in units where the AdS radius is set
to 1)
\be
S_{D3} = -\frac{N}{2\pi^2}\int d^4\s \left( \sqrt{-g} - P[C_4] \right), 
\ee
where $g_{ab} = \p_a X^M \p_b X_M$, where $a,b=0,\ldots,3$ label
the worldvolume coordinates and where $X^M$ are the embedding
coordinates. Note that there is no B-field in our background, and we
also will not be turning on worldvolume gauge fields. The four-form
potential $C_4$ which will be important for the giant graviton has its
legs entirely in the $S^5$, and may be taken as \cite{Grisaru:2000zn}
\be
C_{\phi \chi_1 \chi_2 \chi_3} =\, \cos^4\theta\, \text{Vol}(\O_3),
\ee
where the $\chi_i$ are angles covering the $S^3 \subset S^5$ and where
$\text{Vol}(\O_3)$ indicates its volume element. 

One takes the ansatz
\be \label{ansatz}
\r = 0,\quad \s^0 = t,\quad \phi = \phi(t),\quad \s^i = \chi_i,
\ee
and obtains 
\be \label{lag}
S =\int dt\, L = -N \int dt \,\Bigl[ \cos^3\theta \sqrt{1 -
  \dot\phi^2\sin^2\theta} - \dot\phi \,\cos^4\theta\Bigr].
\ee
Independence of $\phi$ leads to a conserved angular momentum
\be\label{L}
k \equiv \frac{\d  L}{\d \dot\phi} 
= \frac{N\dot\phi \sin^2\theta\cos^3\theta}{ \sqrt{1 -
  \dot\phi^2\sin^2\theta}} + N \cos^4\theta.
\ee
The action may be rewritten in terms of $k$, to give
\be \label{act}
S = N \int dt \,\frac{\cos^4\theta}{\sin\theta} \frac{
    l - \cos^2\theta}{\sqrt{( l-\cos^4\theta)^2 + \sin^2\theta\cos^6\theta}},
\ee
where $l \equiv k/N$. One may also introduce an energy defined by
\be
E \equiv \dot\phi k -  L 
= \frac{N}{\sin\theta}\sqrt{(l-\cos^4\theta)^2+\sin^2\theta\cos^6\theta},
\ee
and which notably removes the WZ part of the action. The energy is
minimized by
\be
\cos^2\theta = l, \qquad E_{\text{min.}} = k, \qquad S_{\text{min.}} = 0,
\label{Emin}
\ee
and by plugging this value in to (\ref{L}), one finds that
\be \label{phi}
\dot \phi = 1.
\ee

\subsection{Giant graviton on $AdS_5$}

We turn next to the giant graviton \cite{Grisaru:2000zn} with
worldvolume $\bR\times S^3 (\subset AdS_5)$. We begin in Lorentzian
signature $(-,+,\ldots,+)$. The metric of $AdS_5\times S^5$ can be
taken as in (\ref{global}). The action for the
anti-D3-brane\footnote{It is the anti-D3-brane which is dual to the
  large symmetric representation gauge theory operator \cite{Grisaru:2000zn}.} is (in
units where the AdS radius is set to 1)
\be
S_{D3} = -\frac{N}{2\pi^2}\int d^4\s \left( \sqrt{-g} + P[C_4] \right), 
\label{D3action}
\ee
where $g_{ab} = \p_a X^M \p_b X_M$, where $a,b=0,\ldots,3$ label the
worldvolume coordinates and where $X^M$ are the embedding
coordinates. The four-form potential $C_4$ which will be important for
this giant graviton has its legs entirely in the $AdS_5$, and may be
taken as \cite{Grisaru:2000zn}
\be\label{CAdS}
C_{t \wt\chi_1 \wt\chi_2 \wt\chi_3} =\, -\sinh^4\r\, \text{Vol}(\wt\O_3),
\ee
where the $\wt\chi_i$ are angles covering the $S^3 \subset AdS_5$ and where
$\text{Vol}(\wt\O_3)$ indicates its volume element. 

One takes the ansatz
\be \label{ansatzAdS}
\r = \text{const.},\quad \s^0 = t,\quad \s^i = \wt\chi_i,
\quad \phi = \phi(t),\quad\theta=\frac{\pi}{2},
\ee
and obtains 
\be \label{lagAdS}
S =\int dt\, L = -N \int dt \,\Bigl[ \sinh^3\r \sqrt{\cosh^2\r -
  \dot\phi^2} - \sinh^4\r\Bigr].
\ee
Independence of $\phi$ leads to a conserved angular momentum
\be\label{LAdS}
\wt k \equiv \frac{\d  L}{\d \dot\phi} 
= \frac{N\dot\phi \sinh^3\r}{ \sqrt{\cosh^2\r -
  \dot\phi^2}}.
\ee
The action may be rewritten in terms of $\wt k$, to give
\be 
S = -N \int dt\cosh\r\sinh^4\r
\left[\sinh^2\r\sqrt{\frac{1}{\sinh^6\r+\wt l^2}} - 1\right],
\ee
where $\wt l \equiv \wt k/N$. One may also introduce an energy defined by
\be
E \equiv \dot\phi \wt k -  L 
= N\left[\cosh\r \sqrt{\sinh^6\r+\wt l^2}-\sinh^4\r\right].
\ee
The energy is minimized by
\be
\sinh^2\r = \wt l, \qquad E_{\text{min.}} = \wt k, \qquad S_{\text{min.}} = 0,
\ee
and by plugging this value in to (\ref{LAdS}), one finds that
\be 
\dot \phi = 1.
\ee

\subsection{Coordinates}
\label{sec:coords}

We can map the global coordinates (\ref{global}) of section
\ref{sec:s5} into the
Poincar\'e patch as follows. Take as a simplification $AdS_3$, for
which the factor $d\wt\O_3^2 = d\psi^2$, then we have that
\bsp 
&z =\frac{R}{  \cosh \r \cos t - \sinh \r \cos \psi},\\ 
&x^0 = \frac{R \cosh \r \sin t}{\cosh \r
\cos t - \sinh \r \cos \psi }, \qquad x^1 = \frac{ R\sinh\r \sin
\psi}{\cosh \r \cos t - \sinh \r \cos \psi },
\end{split}
\ee
where the metric of the Poincar\'e patch is
\be
ds^2 = \frac{-(dx^0)^2+(dx^1)^2 + dz^2}{z^2}.
\ee
On the path of the $S^5$ giant graviton we have $\r = 0$. Continuing to
Euclidean $AdS$, so that $t \to t_E = -i t$ and $x^0 \to x^0_E =- ix^0$ we have that
\be \label{geod}
z = \frac{R}{\cosh t_E},\quad x^0_E = R\tanh t_E, \quad x^1=0
\ee
which gives the trajectory of \cite{Janik:2010gc}, if we identify the Euclidean
time direction in the Poincar\'e patch with the spatial direction
in which the operators are separated on the boundary. Note that the
operator separation is given by $L = 2 R$ \cite{Janik:2010gc}.

In the case of the $AdS_5$ giant graviton we must use the generalization
of the coordinate transformation to $AdS_5$
\bsp 
&z =\frac{R}{ \cosh \r \cos t -   n_0\sinh \r},\\ 
&x^0 = \frac{R \cosh \r \sin t}{\cosh \r
\cos t -  n_0\sinh \r  }, \qquad 
\vec x = \frac{R\, \vec n \sinh\r}{\cosh \r \cos t -  n_0\sinh \r },
\end{split}
\ee
where the $S^3 \subset AdS_5$ is given by the embedding coordinates
$n_I = (n_0,\vec n)$, $n_I n_I=1$.

We remind the reader of the Euclidean form of the D-brane
action\footnote{The anti-D-brane action has a flipped sign on the WZ part.}
\be
S^E_{D3} = \frac{N}{2\pi^2}\int d^4\s \left( \sqrt{g} -i P[C_4] \right), 
\ee
and note that the four-form potential with legs in the $AdS_5$ part of
the geometry (\ref{CAdS}) gains a $-i$ under the Wick rotation, $C_4^{AdS}\to
-iC_4^{AdS}$, due to having a leg in the temporal direction; the
potential on $S^5$ is unaffected. Plugging the
Wick-rotated solutions into the Euclidean action always yields a real result,
since the angle $\phi = -i t$ compensates for the factor of $i$ in the
Wess-Zumino term for the giant graviton on $S^5$, whose four-form
potential has a leg in the $\phi$ direction.
%

\subsection{Supergravity fluctuations}

The supergravity modes that we are interested in are fluctuations of
the 4-form potentials, as well as the spacetime metric, and are dual
to chiral primary operators with R-charge $\D$ in ${\cal N}=4$ SYM 
\cite{Berenstein:1998ij}\cite{Lee:1998bxa}\cite{Kim:1985ez}.
The fluctuations are\footnote{The traceless symmetric double covariant
derivative is defined as $ \nabla_{(\m} \nabla_{\n)} \equiv
\frac{1}{2} \left( \nabla_\m \nabla_\n + \nabla_\n \nabla_\m \right) -
\frac{1}{5} g_{\m \n} \, g^{\r \s} \nabla_\r \nabla_\s$.}
\bsp\label{fluct} 
&\d g_{\m \n} =
\left[-\frac{6\,\D}{5}\,g_{\m \n} + \frac{4}{\D+1} \, \nabla_{(\m}
  \nabla_{\n)} \right] \,s^\D(X)\,Y_\D(\O), \\
&\d g_{\a\b } = 2\,\D\,g_{\a\b
} \,s^\D(X)\,Y_\D(\O),\\
&\delta
C_{\mu_1\mu_2\mu_3\mu_4}=-4\,\e_{\mu_1\mu_2\mu_3\mu_4\mu_5}\nabla^{\mu_5}
\,s^\D(X)\,Y_\D(\O),\\
&\delta C_{\alpha_1\alpha_2\alpha_3\alpha_4}=4\e_{\alpha
  \alpha_1\alpha_2\alpha_3\alpha_4}s^\D(X)\nabla^{\alpha}Y_\D\left(\Omega\right),
\end{split}
\ee
where $\m,\n$ are $AdS_5$ and $\a,\b$
are $S^5$ indices. The symbol $X$ indicates coordinates on $AdS_5$ and
$\O$ coordinates on the $S^5$. The $Y_\D(\O)$ are
the spherical harmonics on the five-sphere, while $s^\D(X)$ have
arbitrary profile and represent a scalar field propagating on $AdS_5$
space with mass squared $=\D(\D-4)$, where $\D$ labels the representation
of $SO(6)$ and must be an integer greater than or equal to 2.

The bulk-to-boundary propagator for $s^\D$ is given in
\cite{Berenstein:1998ij}, with normalization from \cite{Lee:1998bxa}. It is
\begin{equation}\label{prop}
\begin{split}
\sqrt{\frac{\a_0}{B_\D}} \frac{z^\D}{((x-x_B)^2+z^2)^\D} 
\simeq \sqrt{\frac{\a_0}{B_\D}} \frac{z^\D}{x_B^{2\D}},
\end{split}
\ee
where we have indicated the limit where the boundary insertion $x_B^\m$ is
taken infinitely far away from the giant graviton; this is the limit
we will be interested in. The normalization is given by
\begin{equation}
\begin{split}
\alpha_0 = \frac{\D-1}{2\pi^2}, \qquad &B_\D = \frac{2^{3-\D} N^2
  \D(\D-1)}{\pi^2(\D+1)^2}.
\end{split}
\end{equation}
%

\subsection{Antisymmetric giant graviton}
\label{as}

We consider the coupling of the supergravity fluctuations to the Euclidean
action. The DBI part, given by 
\be S_{DBI}=\frac{N}{2\pi^2}\int
d^4\sigma\sqrt{g} ,
\ee
gives the following variation
\be \delta
S_{DBI}=\frac{N}{2}\cos^2\theta\int dt\, Y_\D \left(\Omega\right)
\left(\frac{4}{\Delta+1}\partial^2_t 
-\frac{2\Delta\left(\Delta-1\right)}{\Delta+1}-8\Delta\,\sin^2\theta+6\Delta\right)
s^\Delta .
\ee
We will be interested in the spherical harmonic
\be
Y_\D(\O) = \frac{\sin^\D\theta \,e^{i\D\phi}}{2^{\D/2}} =
\frac{\sin^\D\theta \,e^{\D t }}{2^{\D/2}},
\ee
which corresponds to the gauge theory operator $\Tr Z^\D$.
Replacing the field $s^\Delta$ with the bulk to boundary
propagator (\ref{prop}), namely
\be s^\Delta \to \frac{\Delta+1}{N
\Delta^{\frac{1}{2}}2^{2-\frac{\Delta}{2}}}\frac{z^\Delta}{x_B^{2\D}} ,
\ee
we obtain
\be \label{D3} 
\delta S_{DBI}= \frac{\cos^2\theta
\sin^\D \theta
\left(\Delta+1\right)\sqrt{\Delta} }{2}\int dt\,\frac{{\cal R}^\D\,e^{\Delta
t}}{\cosh^\Delta t}\left(2\,\cos^2\theta-\frac{1}{\cosh^2 t }\right),
\ee
where ${\cal R}\equiv R/x^2_B$, see section \ref{sec:coords}.
We now turn our attention to the Wess-Zumino coupling.
Because the relevant legs of $C_4$ are in $S_5$ we
require only the fluctuation $\delta C_{\phi\chi_1\chi_2\chi_3}$
\be
\delta C_{\phi\chi_1\chi_2\chi_3}=4\e_{\theta \phi\chi_1
\chi_2\chi_3}s^\D\nabla^{\theta}Y_\D\left(\Omega\right)
=2^{-\frac{\Delta}{2}+2}\e_{\theta \phi\chi_1\chi_2\chi_3}
\Delta s^\D \left( \sin\theta\right)^{\Delta -1}\cos \theta 
e^{\Delta  t }.
\ee
Therefore the variation of the Wess-Zumino part is
\bsp \label{wz}
\delta S_{WZ}&= -2^{-\frac{\Delta}{2}+2}N \Delta\int dt\,  e^{\Delta  t }
\sin^\D \theta
\cos^4\theta s^\Delta\\
&=-\cos^4\theta \sin^\D \theta 
\left(\Delta+1\right)\sqrt{\Delta}\int dt\, \frac{{\cal R}^\D e^{\Delta  t }}{\cosh^\Delta t }.
\end{split}
\ee
Adding the variations of the DBI and Wess-Zumino terms we find 
\be \label{SE}
\delta S =-\frac{\cos^2\theta \left(\sin\theta\right)^\Delta
  \left(\Delta+1\right)\sqrt{\Delta}}{2}
\int_{-\infty}^\infty d t\,\frac{{\cal R}^\D e^{\Delta
     t }}{\cosh^{\Delta+2} t }
=-(2{\cal R})^\Delta \sqrt{\Delta} \cos^2\theta \sin^\Delta\theta ,
\ee
or (recalling that $\cos^2\theta = k/N$, $\D=J$) in terms of gauge theory quantities, the three-point function
structure constant is given by
\be
C^A_{k,k-J,J} =  \sqrt{J} \frac{k}{N} \left(1-\frac{k}{N}\right)^{J/2}.
\ee

\subsection{Symmetric giant graviton}
\label{ss}

We write the metric on $S^3 \subset AdS_5$ as
\be
d\wt\O_3^2 = d\vt^2 + \cos^2\vt d\phi_1^2 + \sin^2\vt d\phi_2^2,
\ee
so that embedding coordinates are given by
\be
n_I = (\cos\vt\sin\phi_1,\cos\vt\cos\phi_1,\sin\vt\sin\phi_2,\sin\vt\cos\phi_2).
\ee
The variation of the Lagrangian density is
\bsp
\d {\cal L} = &\frac{N}{4\pi^2}\sinh^2\r \cos\vt\sin\vt\Biggl[-2\D s +
  h_{tt} + h_{\vt\vt} + \frac{h_{\phi_1\phi_1}}{\cos^2\vt}+
  \frac{h_{\phi_2\phi_2}}{\sin^2\vt}\Biggr]\\
&-\frac{2N}{\pi^2}\cosh\r\sinh^3\r \cos\vt \,\p_\r s.
\end{split}
\ee
where the second line is the WZ part of the variation, and
\be
h_{\m\n} = \frac{2}{\D+1}\Bigl[2\nabla_\m\nabla_\n-\D(\D-1)g_{\m\n}\Bigr]s,
\ee
where $s=s^\D Y_\D$, while
\bsp
&\nabla_t\nabla_t s = \left(\p_t^2 + \cosh\r\sinh\r\,\p_\r\right)s,\\
&\nabla_\vt\nabla_\vt s = \bigl(\p_\vt^2 + \cosh\r\sinh\r\,\p_\r\bigr)s,\\
&\nabla_{\phi_1}\nabla_{\phi_1} s = \left(\p_{\phi_1}^2 +
\cos^2\vt\cosh\r\sinh\r\,\p_\r-\cos\vt\sin\vt\,\p_\vt\right)s,\\
&\nabla_{\phi_2}\nabla_{\phi_2} s = \bigl(\p_{\phi_2}^2 +
\sin^2\vt\cosh\r\sinh\r\,\p_\r+\cos\vt\sin\vt\,\p_\vt\bigr)s.
\end{split}
\ee
Now we may replace the field $s$ with the bulk-to-boundary propagator (\ref{prop})
\be
s\to
\frac{\D+1}{2^{2}\sqrt{\D}N} \frac{{\cal R}^\D e^{\D t}}{ \left(\cosh
  \r \cosh t -  \cos\vt\sin\phi_1\sinh \r\right)^\D},
\ee
where ${\cal R}\equiv R/x^2_B$, see section \ref{sec:coords}.
There is a great simplification which occurs between the DBI and WZ
pieces of the variation of the action, which leads to
\bsp
\d S= &-\int_{-\infty}^{\infty} dt \int_0^{2\pi}d\phi_1
\int_0^{2\pi}d\phi_2 \int_0^{\pi/2}d\vt \\
&\times \frac{\sqrt{\D}(\D+1)}{4\pi^2}
\cos\vt\sin\vt\sinh^2\r\frac{{\cal R}^\D e^{\D t}}{\left(\cosh \r \cosh t -
\cos\vt\sin\phi_1\sinh \r\right)^{\D+2}} ,
\end{split}
\ee
where we have included the spherical harmonic $Y=e^{\D
  t}/2^{\D/2}$. We may re-cast the integral as follows 
\bsp\label{theint}
&\d S = -\frac{\sqrt{\D}(\D+1)}{2\pi} \frac{\sinh^2\r}{\cosh^{\D+2}\r}
\int_{-\infty}^{\infty} dt\int_0^{2\pi}d\phi_1\int_0^1d\l
\frac{{\cal R}^\D e^{\D
    t}}{\cosh^{\D+2}t}\frac{\l}{\left[1-\frac{\l\sin\phi_1\tanh\r}{\cosh t}\right]^{\D+2}}\\
& =-\frac{\sqrt{\D}(\D+1)}{2\pi} \frac{\sinh^2\r}{\cosh^{\D+2}\r}
\int_{-\infty}^{\infty} dt\int_0^{2\pi}d\phi_1\int_0^1d\l
\frac{{\cal R}^\D e^{\D
    t}}{\cosh^{\D+2}t}\\
&\qquad\qquad\qquad\qquad\qquad\qquad\qquad\qquad\times
\l \sum_{k=0}^\infty \left(\frac{\l\sin\phi_1\tanh\r}{\cosh
  t}\right)^k\frac{\G(\D+k+2)}{\G(k+1)\G(\D+2)}\\
&=-\frac{\sqrt{\D}}{2\pi\G(\D+1)} \frac{\sinh^2\r}{\cosh^{\D+2}\r}
\int_{-\infty}^{\infty} dt\int_0^{2\pi}d\phi_1
\frac{{\cal R}^\D e^{\D
    t}}{\cosh^{\D+2}t}\\
&\qquad\qquad\qquad\qquad\qquad\qquad\qquad\times
 \sum_{k=0}^\infty \frac{1}{k+2}\left(\frac{\sin\phi_1\tanh\r}{\cosh
  t}\right)^k\frac{\G(\D+k+2)}{\G(k+1)}\\
&=-\frac{\sqrt{\D}}{2\G(\D+1)} \frac{1}{\cosh^{\D}\r}
\sum_{k=0}^\infty\int_{-\infty}^{\infty} dt
\frac{{\cal R}^\D e^{\D
    t}}{\cosh^{\D+2+2k}t}
  \frac{1}{2^{2k}}\frac{\G(\D+2k+2)}{\G(k+2)\G(k+1)}
\tanh^{2k+2}\r \\
&=-\frac{(2{\cal R})^\D}{\sqrt{\D}\G(\D)} \frac{1}{\cosh^\D\r}\sum_{k=0}^\infty
\tanh^{2k+2}\r \frac{\G(\D+k+1)}{\G(k+2)}\\
&=-\frac{(2{\cal R})^\D}{\sqrt{\D}} \Bigl( \cosh^\D\r - \cosh^{-\D}\r \Bigr).
\end{split}
\ee
In terms of gauge theory quantities, recalling that $
\sinh^2\r = k/N$, $\D=J$, the three-point function
structure constant is given by
\be
C^{S}_{k,k-J,J} =\frac{1}{\sqrt{J}} \Biggl( \left(1+\frac{ k}{N}\right)^{J/2}
-  \left(1+\frac{ k}{N}\right)^{-J/2} \Biggr).
\ee

\section{Discussion and Conclusion}

We have obtained the three-point function involving two giant
gravitons and one pointlike graviton from the gauge theory as well as
from the string theory side. On the string theory side the calculation
was carried out in a semiclassical approximation where the two giant
gravitons were heavy and the pointlike one light.  Both the giant
gravitons and the pointlike graviton were moving with given angular
velocities on $S^5$.  In the case where the
two giant gravitons were wrapping an $S^3\subset AdS_5$ we found for
the three-point function 
\beq 
C^{S, string}_{k,k-J,J}
=\frac{1}{\sqrt{J}} \Biggl( \left(1+\frac{ k}{N}\right)^{J/2} -
\left(1+\frac{ k}{N}\right)^{-J/2} \Biggr).  
\eeq 
Here $k$
is the $S^5$ angular momenta of the giant gravitons and $J$ the $S^5$
angular momentum of the pointlike graviton.  First, we notice that in
the limit where the size of the giant gravitons shrinks to zero, i.e.\
$\frac{k}{N}\rightarrow 0$, we recover, as we should, the result for
the three point function of three pointlike gravitons, cf.\
eqn.~\rf{3ptchiral} 
\beq 
C_{k,k-J,J}^{S, string} \rightarrow
\frac{\sqrt{J}\, k}{N},\hspace{0.7cm} \mbox{for}
\hspace{0.7cm}\frac{k}{N}\rightarrow 0.  
\eeq 
Secondly, we observe
that as the (unrestricted) angular momentum of the giant gravitons
becomes large, i.e.\ $\frac{k}{N}\rightarrow \infty$ the three-point
function turns into the gauge theory three-point function involving
two Schur polynomials and one chiral primary, cf.\ eqn~\rf{CS} 
\beq
C_{k,k-J,J}^{S, string} \rightarrow \frac{1}{\sqrt{J}} \left(1+\frac{
k}{N}\right)^{J/2} =\,\,\, C_{k,k-J,J}^{S, gauge} \hspace{0.7cm}
\mbox{for} \hspace{0.7cm}\frac{k}{N}\rightarrow \infty.  
\eeq 
Although a more natural parameter region
to consider would be $\frac{k}{N}\sim 1$, this matching at large $k$ is an interesting 
coincidence. It may point to an easing, in the symmetric case, of the problems of the Schur
polynomial description of giant gravitons discussed below.

In the case where the giant graviton was wrapping an $S^3\subset S^5$ 
we found from the string theory calculations
\beq
C^{A, string}_{k,k-J,J}
=\sqrt{J} \,\frac{k}{N} \left(1-\frac{ k}{N}\right)^{J/2} . 
\eeq
Again, we notice that we correctly 
recover the three-point function of three point like
gravitons in the limit where the size of the giant gravitons shrink to 
zero, i.e.
\beq 
C_{k,k-J,J}^{A, string} \rightarrow
\frac{\sqrt{J}\, k}{N},\hspace{0.7cm} \mbox{for}
\hspace{0.7cm}\frac{k}{N}\rightarrow 0.  
\eeq
The gauge theory analysis gave the result
\beq
C^{A, gauge}_{k,k-J,J}
=(-1)^{(J-1)}\frac{1}{\sqrt{J}} \, \left(1-\frac{ k}{N}\right)^{J/2}.  
\eeq
In this case the giant graviton angular momentum $k$ has to
satisfy the bound  of $\frac{k}{N}\leq 1$. For $\frac{k}{N}=1$ we find
that both the string theory and gauge theory result are exactly equal to zero
but considering the limit $\frac{k}{N}\rightarrow 1$ the two results differ
by a factor proportional to  $J$.\footnote{As mentioned previously, the $(-1)^{J}$ part of the prefactor 
is convention dependent and could be removed by a different normalisation.} 

Our interpretation of the mismatch between the gauge and string theory
calculations centers on the validity of the
Schur polynomials as duals of the giant gravitons. It is well known
that the Schur polynomials should only describe the giant gravitons
when the size of the operators $k$ is of order $N$, in the large-$N$
limit \cite{Balasubramanian:2001nh}. Furthermore, the Schur polynomials do not reduce to chiral
primaries in the small-$\frac{k}{N}$ limit, and are therefore disconnected from
the pointlike limit of the giant gravitons.  When calculating a
three-point function with a small operator such as the pointlike
graviton dual, i.e. the chiral primary $\Tr Z^J$, we are probing the
chiral-primary content of the OPE of two Schur polynomials $\chi_k(\bar{Z})$
and $\chi_{k-J}(Z)$, schematically
\be
 \chi_k(\bar{Z}(0))\, \chi_{k-J}(Z(x)) = \ldots + C^{gauge}_{k,k-J,J} 
\Tr \bar{Z}^J (0)\,
x^{-2J} + \ldots.
\ee
We can certainly trust the terms in the OPE involving operators of
dimension ${\cal O}(N)$, but it is not clear that the small-operator
content of the operators $\chi_k(Z)$, i.e. the structure constants $
C^{gauge}_{k,k-J,J}$, are in fact themselves dual to the three-point
functions of two giant gravitons and one pointlike graviton, defined
holographically in string theory. Indeed we find that the string
theory results interpolate smoothly to the limit where all three
gravitons are pointlike, whereas, without surprise, the gauge theory
results fail to do so.

The fact that for the symmetric case we get a string three-point function which
nicely interpolates between the gauge theory three-point function of
single trace chiral primaries and that of Schur polynomials raises the
question of whether in the gauge theory language one can construct an
(orthogonal) basis of operators which interpolates between single
trace operators and the Schur polynomials in such a way that the gauge
theory and string theory three-point functions can be exactly matched
for all values of $\frac{k}{N}$.  The question of the existence of
such an interpolating basis has been brought up before, see
e.g. \cite{Balasubramanian:2001nh,Dhar:2005su}, but still lacks 
resolution. The string theoretic results given here will hopefully serve as a
benchmark for any such future construction.

Although we have not emphasized the point, it is also possible that there exist
subtleties in the generalization of the methods developed for the holographic
two-point functions involving semi-classical strings in \cite{Janik:2010gc} to the
case of D-branes, ultimately affecting the procedure used here to calculate
holographic three-point functions. For this reason, a first principles derivation
using D-branes, as was presented in \cite{Janik:2010gc}
for strings, would be welcome. We make some progress toward this goal in
appendix \ref{2ptapp} where the holographic two-point function for giant gravitons on
$S^5$ is derived.

Finally, let us mention that this study, while being concerned only
with 1/2 BPS objects, opens up the avenue for studying holographic
three-point functions of a variety of extended objects in the form of
branes, an interesting example being the systems of non-BPS giant
gravitons moving with two angular momenta on $S^5$ and being dual to
so-called restricted Schur polynomials, see \cite{Carlson:2011hy} and
references therein.

\section*{Acknowledgements}

We thank I.\ Kostov, D.\ Serban and K.\ Zarembo for discussions.
CK, DY and KZ were supported by FNU through grant number 272-08-0329. AB
was supported in part by the Niels Bohr International Academy.

\appendix 

\section{ A shortcut to the $S^5$ giant graviton two-point function} \label{2ptapp}

 In this appendix we show that the calculation of the holographic $S^5$ graviton two-point
function can easily be performed by making use of the fact that,  
as noted in \cite{Caldarelli:2004yk}, all the dependence on the $S^5$
directions can be integrated out, leaving a point particle on $AdS_5$ with effective mass equal
to the dimension of the giant graviton.

 Let us start with the DBI action (\ref{D3action}) for the giant graviton and take the embedding
to be $\sigma^0=t, \sigma^i=\chi_i, \phi=\phi(t)$ but leave the AdS part general. Integrating over
the three-sphere coordinates we find
\be 
S_{D3}=-N\cos^3\theta\int\diff t\sqrt{-G_{MN}\dot{X}^M\dot{X}^N}+N \cos^4\theta \int\dot{\phi}\diff t \;.
\ee
Now, following \cite{Caldarelli:2004yk}, we switch to a Polyakov-type formulation, by introducing an einbein $e$:
\be
S_{D3}=\half \int\diff t\left(\frac1e G_{MN}\dot{X}^M\dot{X}^N-m^2 e\right)+N\cos^4\theta\int\dot{\phi}\diff\sigma_0,
\ee
where we have defined an effective mass, 
\be
m=N \cos^3\theta\;.
\ee
To go back to the original action one simply solves for $e=\frac1m\sqrt{-G_{MN} \dot{X}^M\dot{X}^N}$ 
and substitutes back. Now let us separate the $AdS_5$ from the $S^5$ part as
\be
G_{MN}\dot{X}^M\dot{X}^N=g_{\mu\nu} \dot{x}^\mu\dot{x}^\nu+\sin^2\theta~ \dot\phi^2,
\ee
to obtain
\be \label{action}
S=\half\int\diff t \left(g_{\mu\nu} \dot x^\mu\dot x^\nu-m^2 e+\frac{1}{e}\sin^2\theta {\dot\phi}^2
+2N\cos^4\theta\dot\phi\right) \;.
\ee
The conjugate momentum to $\phi$ is
\be \label{conjmom}
k=\frac{1}{e}\dot\phi\sin^2\theta+N\cos^4\theta,
\ee
which of course agrees with (\ref{L}) if we substitute the on-shell value of $e$ (together with
the form of the metric (\ref{global}) and $\rho=0$).  
Now fix $\theta$ to the value $\theta_0$ minimising the energy (\ref{Emin})
\be
\cos^2\theta_0=l=\frac{k}{N}\;.
\ee
Substituting $k$ from (\ref{conjmom}) and solving for $\dot\phi$ we get
\be
\dot\phi=e N\cos^2\theta_0\;.
\ee
It is easy to see that this would lead to (\ref{phi}) were we to use the on-shell value of the einbein.  
Substituting $\dot\phi$ into (\ref{action}) we finally find
\be
-m^2 e+\frac{1}{e}\sin^2\theta_0 {\dot\phi}^2+2N\cos^4\theta_0\dot\phi =e~ N^2\cos^4\theta_0=e~M_{D3}^2\;,
\ee
to conclude \cite{Caldarelli:2004yk} that the giant graviton can be described by a particle moving in 
$AdS_5$ with mass:
\be
M_{D3}= N\cos^2\theta_0=k\;.
\ee
Standard point-particle techniques (reviewed in \cite{Janik:2010gc}) now straightforwardly give for the
two-point function
\be
G(0,\epsilon;x,\epsilon)=\left(\frac{|x|}{\epsilon}\right)^{-2M_{D3}}
=\left(\frac{|x|}{\epsilon}\right)^{-2k},
\ee
which is the expected answer.\footnote{This result was also recently obtained in \cite{Bak:2011yy}
via an alternative method.}


\bibliography{ggwriteup2}%
\end{document}